\documentclass[twoside]{article}
\usepackage{fleqn,espcrc2}
\usepackage{graphicx}

\newcommand{\gbare}{g^{2}_{0}}
\newcommand{\csw}{c_{\mathrm{sw}}}
\def\simlt{\rlap{\lower 3.5 pt\hbox{$\mathchar \sim$}}\raise 1pt \hbox {$<$}}
\def\simgt{\rlap{\lower 3.5 pt\hbox{$\mathchar \sim$}}\raise 1pt \hbox {$>$}}

\title{
   \vspace*{-2.0em}
   {\normalsize \hfill {\sf KEK-CP-145, UT-CCP-P139, UTHEP-473}\\ }
   \vspace*{1.0em}
Study of finite volume effects in the non-perturbative 
determination of $\csw$ with the SF method in full 
three-flavor lattice QCD\thanks{Presented by K-I.I.}
}

\newcommand{\Tsukuba}%
{Institute of Physics, University of Tsukuba, Tsukuba, Ibaraki 305-8571, Japan}

\newcommand{\RCCP}%
{Center for Computational Physics, University of Tsukuba, Tsukuba, Ibaraki 305-8577, Japan}

\newcommand{\ICRR}%
{Institute for Cosmic Ray Research, University of Tokyo, Kashiwa, Chiba 277-8582, Japan}

\newcommand{\KEK}%
{High Energy Accelerator Research Organization (KEK), Tsukuba, Ibaraki 305-0801, Japan}

\newcommand{\YITP}%
{Yukawa Institute for Theoretical Physics, Kyoto University, Kyoto 606-8502, Japan}

\newcommand{\Hiroshima}%
{Department of Physics, Hiroshima University, Higashi-Hiroshima, Hiroshima 739-8526, Japan}

\newcommand{\RBRC}%
{RIKEN BNL Research Center, Brookhaven National Laboratory, Upton, NY, 11973, USA}
       
\author{CP-PACS and JLQCD Collaborations:
K-I.~Ishikawa\rlap,\address{\Tsukuba}$^{,}$\address{\RCCP}
S.~Aoki\rlap,$^{\rm a}$ 
M.~Fukugita\rlap,\address{\ICRR}
S.~Hashimoto\rlap,\address{\KEK}
N.~Ishizuka\rlap,$^{\rm a,b}$
Y.~Iwasaki\rlap,$^{\rm a}$
K.~Kanaya\rlap,$^{\rm a}$
T.~Kaneko\rlap,$^{\rm d}$
Y.~Kuramashi\rlap,$^{\rm d}$
V.~Lesk\rlap,$^{\rm b}$
M.~Okawa\rlap,\address{\Hiroshima}
N.~Tsutsui\rlap,$^{\rm d}$
A.~Ukawa\rlap,$^{\rm a,b}$
T.~Umeda\rlap,\address{\YITP}
N.~Yamada\rlap,\address{\RBRC}
T.~Yoshi\'e$^{\rm a,b}$
}

\begin{document}

\begin{abstract}
The non-perturbative $\csw$ determined 
by the Schr\"{o}dinger functional (SF) method with the RG-improved gauge 
action in dynamical $N_f=3$ QCD shows a finite volume effect
when the numerical simulations are carried out at a constant lattice size $L/a$.
We remove the unwanted finite volume effect by keeping physical lattice extent $L$ 
at a constant.  The details of the method and the result obtained 
for non-perturbative $\csw$ with a constant $L$ are reported.  
\vspace*{-1.3em}
\end{abstract}

\maketitle

\section{Introduction}
\label{sec:intro}
\vspace*{-0.0em}
Last year we reported the existence of a sizable finite volume 
effect in the non-perturbative $\csw$ determined by the Schr\"{o}dinger functional (SF)
method at a fixed lattice size ($L/a$$=$const) when the RG-improved gauge action 
is adopted in full three-flavor QCD~\cite{CPPACS_JLQCD_NPT_CSW}.
This effect causes a constant deviation from the true non-perturbative
value of $\csw$, and yields $O(a)$ errors in physical observables.

This year we explain our strategy for removing the unwanted
finite volume effect from $\csw$ and report
the bare coupling dependence of the non-perturbative $\csw$ 
for three-flavor dynamical QCD. 

\vspace*{-0.5em}
\section{Method and Simulations}
\label{sec2}

The basic strategy
to obtain the non-perturbative $\csw$ 
with the SF setup is
described in~\cite{JS}.
We used Eq.~(2.17) of~\cite{AFW} for the SF boundary coupling of 
the RG-improved gauge action. 
The polynomial Hybrid Monte Carlo (PHMC) algorithm~\cite{JLQCD_PHMC} 
is employed for the numerical simulation of the three-flavor dynamical QCD.

To remove the finite lattice size effect we 
fix the physical lattice extent $L$ so that $a/L$ corrections vanish with $a$.
This can be achieved by the following steps.\\
\noindent
(1) Define the reference lattice size $L^{*}/a$ at a bare coupling 
   value ${\gbare}^{*}$.
   This also defines the reference physical lattice extent $L^{*}$.\\
(2) For other $\gbare$'s, the lattice sizes $L^{*}/a$ at each $\gbare$ 
    are estimated by the beta function.\\
(3) Determine $\csw$ and $\kappa_c$ via the PCAC condition with the SF setup
   at each $\gbare$ on the corresponding lattice size $L^{*}/a$.

In actual numerical simulations we employ
the universal two-loop beta function to estimate $L^{*}/a$.
This causes a systematic error in the non-perturbative $\csw$, 
which could be large in the strong coupling region.
This systematic error, however, can be partly avoided by defining $L^{*}$ at
the largest coupling value.
We can also examine the magnitude of the error by investigating 
the $L/a$ dependences because the error only appears through the 
incorrect value of $L^{*}/a$.

With the above consideration we take the following setup in the simulations.\\
\noindent
(i)   We define $L^{*}$ so that
      $L^{*}/a=6$ holds at $\beta=6/\gbare=1.9$.\\
(ii)  We employ both $L/a=8$ and $6$ at $\beta=2.0$, and $L/a=8$ 
      for other weaker couplings.\\
(iii) We correct $\csw$ and $\kappa_c$ at $L/a$ to those at $L^{*}/a$ 
      using the one-loop correction calculated 
      with the SF setup in finite lattice volume~\cite{AFW}.\\
Although the perturbative error from setup (iii) appearers 
in strong coupling region, the magnitude can be reduced 
when $L^{*}/a \sim L/a$ holds and $L/a$ dependence is small.
The small $L/a$ dependence is confirmed at $\beta=2.0$ as shown later.
We will briefly discuss the possible resolution of the systematic 
error from the use of perturbative beta function in the last section.

We use the following non-perturbative tuning condition for $\csw$ and $\kappa$.
\begin{equation}
  \label{eq:PCAC}  
\begin{array}{l}
       M(\gbare,L/a,\kappa,\csw)=0,\\
\Delta M(\gbare,L/a,\kappa,\csw)=0,
\end{array}
\end{equation}
where $M(\gbare,L/a,\kappa,\csw)$ is the bare PCAC quark mass and
$\Delta M(\gbare,L/a,\kappa,\csw)$ is the difference of the two bare PCAC quark
masses with different definition. The actual definitions are 
explained in~\cite{JS}. 
The difference from the tuning condition we employed last year
is the absence of the tree-level finite PCAC quark mass which arises from 
the SF boundary for a finite lattice extent.
The tree-level finite volume effect is tuned to the 
corresponding value at $L^{*}/a$ by setup (iii) 
together with the one-loop finite volume effect.

As shown in setup (iii), we use the following one-loop perturbative correction
to convert $\csw$ and $\kappa$ at $L/a$ to those at $L^{*}/a$ at
each $\gbare$;
\begin{equation}
  \begin{array}{l}
 \kappa_{c}(L^{*}/a)=
 \kappa_{c}(L/a)   +\delta \kappa_{c}(L/a;L^{*}/a),\\
       \csw(L^{*}/a)=
       \csw(L/a)   +\delta       \csw(L/a;L^{*}/a).
  \end{array}
  \label{eq:NPTCOND}
\end{equation}
$\delta \kappa_{c}(L/a;\!L^{*}/a)$ and
$\delta \csw(L/a;\!L^{*}/a)$ are defined as
\begin{equation}
  \label{eq:PTCorr}
  \begin{array}{l}
 \delta \kappa_{c}(L/a;\!L^{*}/a)\!=\!
                  \left[\kappa_{c}(L^{*}/a)
                   \!-\!\kappa_{c}(L/a)\right]_{\mbox{\tiny 1-loop}},\\
 \delta       \csw(L/a;\!L^{*}/a)\!=\!
                 \left[\csw(L^{*}/a)
                  \!-\!\csw(L/a)\right]_{\mbox{\tiny 1-loop}},
  \end{array}
  \hspace*{-2em}
\end{equation}
where the right hand side of Eq.~(\ref{eq:PTCorr}) is calculated
up to one-loop level.

\vspace*{-0.5em}
\section{Results}
\label{sec3}

\begin{figure}[t]
\vspace*{-0.5em}
\centering\hspace*{-3mm}
\includegraphics[scale=0.635,clip]{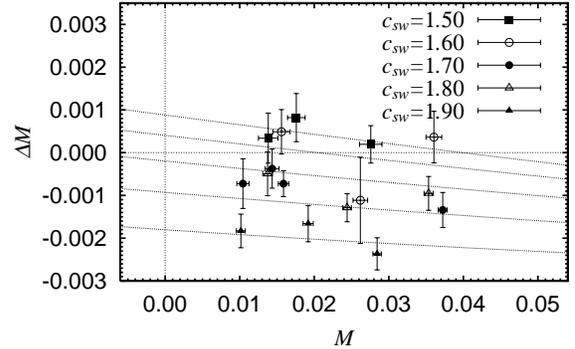}
\vspace*{-4.2em}
\caption{$M$ vs $\Delta M$ at $\beta=2.0$, $L/a=8$.}
\vspace*{-2.5em}
\label{fig:MdM}
\end{figure}

We have carried out the numerical simulation in the region 
$\beta = 12.0-1.9$ with the setup explained in Sec.~\ref{sec2}.
It was found that below $\beta=2.2$ the PHMC
algorithm failed at vanishing or negative PCAC quark masses 
due to large quantum fluctuations. 
We extrapolated them from $M > 0$ to $M = 0$ as shown in 
Fig.~\ref{fig:MdM}. 
It was also found that below $\beta = 1.9$ with $L/a = 8$
it became difficult to satisfy the condition $\Delta M = 0$, 
which is why we employ $L^{*}/a = 6$ at $\beta = 1.9$
for the definition of the reference physical lattice extent.

\begin{figure}[t]
\vspace*{-0.5em}
\centering\hspace*{-3mm}
\includegraphics[scale=0.635,clip]{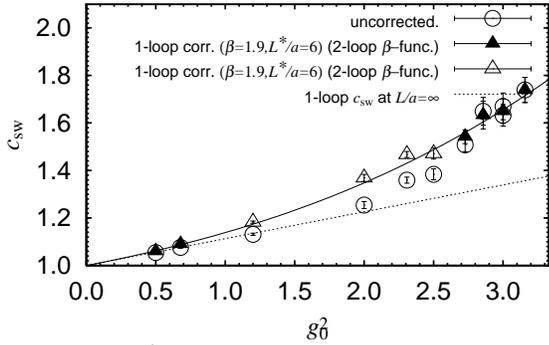} 
\vspace*{-4.3em}
\caption{$\gbare$ dependence of $\csw$. 
         Open circles are the result before $L/a$ tuning.
         Open and filled triangles are tuned to $L^{*}/a$.
         Filled triangles are used for curve fitting.
         Solid line is the fit result.
         Dotted line shows the one-loop result at $L/a$$=$$\infty$.}
\vspace*{-2.4em}
\label{fig2}
\end{figure}

\begin{figure}[t]
\vspace*{-0.5em}
\centering\hspace*{-3mm}
\includegraphics[scale=0.635,clip]{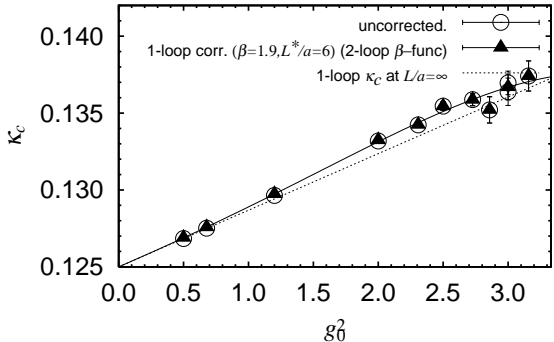}
\vspace*{-4.3em}
\caption{Same as Fig.~\ref{fig2}, but for $\kappa_c$.}
\vspace*{-2.0em}
\label{fig3}
\end{figure}

Figures~\ref{fig2} and \ref{fig3} are the result for $\csw$ and $\kappa_{c}$.
Open circles are the result without $L/a$ matching.
Triangles are after tuning to $L^{*}/a$ by the one-loop correction 
(or interpolation at $\gbare = 3.0$).
For $\csw$ we observe large one-loop corrections in the region 
$1.0 < \gbare < 2.7$.
This is due to rather strong coupling and a large discrepancy 
between $L/a$ and $L^{*}/a$.
We discard the data from the functional fit.
At strong couplings, $\gbare = 6/2.2$ and $6/2.1$, 
the one-loop correction is rather small since $L/a \sim L^{*}/a$. 
The $L/a$ dependence at $\gbare = 3.0$ $(\beta = 2.0)$ is 
negligible compared to the statistical error
(upper circle: $L/a = 8$, lower circle: $L/a = 6$).
We have $6 < L^{*}/a < 8$ via the two-loop beta-function at 
$\gbare = 3.0$, we simply interpolate $\csw$ to $L^{*}/a$ using 
both lattice size data.
While uncorrected data undershoots the one-loop result for 
$L/a = \infty$ (dotted line) 
in the weak coupling region~\cite{CPPACS_JLQCD_NPT_CSW}, 
those with one-loop correction gradually approaches it 
($L^{*}/a$ is essentially infinite in the region).
For $\kappa_c$ the one-loop correction is rather small.
We conclude that the non-perturbative results in the weak coupling region
is consistent with the one-loop one for both $\csw$ and $\kappa_c$.
We fit the filled triangles to obtain the functional form in $\gbare$ for 
$\csw$ and $\kappa_c$. 


After the curve fitting including the knowledge of the 
one-loop perturbative coefficient we obtain
\begin{equation}
  \begin{array}{l}
\displaystyle
\csw(\gbare) = 1 + 0.113 g^{2}_{0}
                 + 0.0209(72) g^{4}_{0}\\
\hspace*{3.4cm}  +\, 0.0047(27) g^{6}_{0},
  \end{array}
\hspace*{-3cm}
\label{NPTC}
\end{equation}
with $\chi^{2}/\mathrm{DOF} = 0.5754$, and
\begin{equation}
  \begin{array}{l}
\displaystyle
\kappa_{c}(\gbare) = 0.125
               +\left(  3.681192 \,\times\! 10^{-3}\right) g^{2}_{0}\\
\hspace*{2.3cm}+\left(  0.141(43)\!\times\! 10^{-3}\right) g^{4}_{0}\\
\hspace*{2.3cm}+\left(  0.124(67)\!\times\! 10^{-3}\right) g^{6}_{0}\\
\hspace*{2.3cm}-\left(  0.049(21)\!\times\! 10^{-3}\right) g^{8}_{0},
  \end{array}
\hspace*{-3cm}
\label{NPTK}
\end{equation}
with $\chi^{2}/\mathrm{DOF} = 1.0014$.
All data are statistically independent
 and $\chi^{2}/\mathrm{DOF}$ 
results in reasonable values.
We show statistical errors only. The systematic error from 
the use of perturbative beta function is not estimated. 
We, however, consider that the systematic error is minimized since
the slope in $\gbare = 3.0-3.15$ of $\csw$ should be
correct because of the negligible $L/a$ dependence at $\gbare = 3.0$.

\vspace*{-0.5em}
\section{Summary}
\label{sec4}

We have extracted the non-perturbative $\csw$ as a function of 
$\gbare$ for $N_f = 3$ with the RG-improved gauge action. 
The functional form is shown in Eq.~(\ref{NPTC}).
In deriving the non-perturbative values at a fixed physical 
lattice extent $L^{*}$,
we used the two-loop beta-function to obtain the corresponding 
lattice size $L^{*}/a$ at a given $\gbare$. 
This may contradict the word ``non-perturbative''.
In order to improve this point we should iteratively execute our procedure 
with the non-perturbative beta-function that is measured with 
the non-perturbative $\csw$ extracted from the previous iteration.
We leave this iterative study for future studies.
The first step to obtain $N_f = 2+1$ hadron mass spectrum has been 
started using the non-perturbative $\csw$ of 
Eq.~(\ref{NPTC})~\cite{JLQCD_CPPACS_REAL}.

This work is supported by Large Scale Simulation Program
No. 98 (FY2003) of High Energy Accelerator Research
Organization (KEK), and also
in part by Grants-in-Aid of the Ministry of Education
(Nos.
13135204,  
14046202,  
15204015,  
15540251,  
15540279). 
N.Y. is supported by the JSPS Research Fellowship.

\end{document}